\documentclass[pra,reprint,amsmath,amssymb,superscriptaddress,longbibli
ography,eqsecnum]{revtex4-2}

\usepackage{graphicx,relsize,epstopdf,color,mathtools,bm,newtxtext,newtxmath,braket,rotating,dsfont,mathtools}

\usepackage[hyphenbreaks]{breakurl}
\usepackage[colorlinks=true,linkcolor=blue,citecolor=blue,urlcolor =blue]{hyperref}

\newcommand{\Tr}{\mathop{\mathrm{Tr}} \nolimits}

\begin{document}

\title{Optimizing measurement tradeoffs in multiparameter spatial superresolution}

\author{J. \v{R}eh\'a\v{c}ek} 
\affiliation{Department of Optics,
 Palack\'y University, 17. listopadu 12, 771 46 Olomouc, 
Czech Republic}

\author{J. L. Romero} 
\affiliation{Departamento de \'Optica,
  Facultad de F\'{\i}sica, Universidad Complutense, 
28040~Madrid,  Spain} 

\author{A. Z. Goldberg}
\affiliation{National Research Council of Canada, 100 Sussex Drive, Ottawa, Ontario K1N 5A2, Canada}
\affiliation{Department of Physics, University of Ottawa, 25 Templeton Street, Ottawa, Ontario, K1N 6N5 Canada}

\author{Z. Hradil}
\affiliation{Department of Optics,
 Palack\'y University, 17. listopadu 12, 771 46 Olomouc, 
Czech Republic}

\author{L. L. S\'{a}nchez-Soto} 
\affiliation{Departamento de \'Optica,
  Facultad de F\'{\i}sica, Universidad Complutense, 
28040~Madrid,  Spain} 
\affiliation{Max-Planck-Institut f\"ur die Physik des Lichts,
  Staudtstra{\ss}e 2, 91058 Erlangen,
  Germany}

\begin{abstract}
The quantum Cram\'er-Rao bound for the joint estimation of the centroid and the separation between two incoherent point sources cannot be saturated. As such, the optimal measurements for extracting maximal information about both at the same time are not known. In this work, we ascertain these optimal measurements for an arbitrary point spread function, in the most relevant regime of a small separation between the sources. Our measurement can be adjusted within a set of tradeoffs, allowing more information to be extracted from the separation or the centroid while ensuring that the total information is the maximum possible.  
\end{abstract}

\maketitle

\section{Introduction}

The Rayleigh criterion~\cite{Rayleigh:1879ab} is a commonly accepted  standard that specifies the minimum separation between two incoherent point sources using a linear imaging system. It was widely believed that this represents a fundamental limitation inherent to the nature of light.  

Recently, Tsang and coworkers~\cite{Tsang:2016aa, Nair:2016aa, Ang:2016aa,Tsang:2017aa} challenged this Rayleigh \emph{curse}.  They resorted to the concepts of quantum Fisher information (QFI) and the associated Cram\'er-Rao  bound (QCRB)~\cite{Helstrom:1976aa} to quantify how well the separation between two point sources can be estimated.  When only the intensity at the image is measured, the Fisher information falls to zero as the separation between the sources decreases and the classical Cram\'er-Rao  bound  (CRLB) diverges accordingly. However, when the QFI for the complete field is calculated, it stays constant and so does the QCRB. This clearly reveals that the Rayleigh limit is not intrinsic to the problem.

Since then, a plethora of publications have arisen from this idea: experimental implementations~\cite{Paur:2016aa,Yang:2016aa,Tham:2017aa,Zhou:2019ab,Tan:2023aa,Santamaria:2024aa}, generalizations~\cite{Rehacek:2017aa,Rehacek:2017ab,Parniak:2018aa,Bisketzi:2019aa,Zhou:2019aa,Napoli:2019aa,Tsang:2019aa,Prakash:2022aa,Matlin:2022aa,Matlin:2022aa}, effects of noise~\cite{Kurdzialek:2022aa,Kurdziaek:2023aa,Gessner:2020aa,Schlichtholz:2024aa} and coherence~\cite{Larson:2018aa,Tsang:2019ab,Hradil:2019aa,Wadood:2021aa,De:2021aa}. A comprehensive review can be found in \cite{Tsang:2019ac} and an updated list of references in \cite{Tsang:aa}.

An optimal method, dubbed spatial mode demultiplexing (SPADE), was designed to estimate the separation between two point sources. This involves passively projecting the collected light onto an orthonormal basis of spatial modes, which is accomplished using a mode-sorting device that couples each spatial mode  to a specific intensity-resolving detector. However, SPADE requires prior knowledge of the centroid of these sources and the alignment of the device with the centroid~\cite{Tsang:2016aa,Prasad:2019aa}. This prior information can be easily obtained because direct imaging is accurate in estimating the centroid. An adaptive two-stage detection scheme that dynamically allocates the resources between the centroid estimation and the separation estimation has been proposed~\cite{Grace:2020aa}, demonstrating that it still outperforms direct imaging. The impact of misalignment on separation estimation has been alsoconsidered~\cite{Almeida:2021aa}. 

At a fundamental level, this is a multiparameter problem. It is well-known that in this scenario the QCRB suffers from the incompatibility issue of quantum measurements~\cite{Zhu:2015wt,Heinosaari:2016wb,Ragy:2016aa,Szczykulska:2016aa,Sidhu:2020aa,Liu:2020aa,Goldberg:2021aa}. It has been noticed~\cite{Chrostowski:2017aa} that the optimal measurements for individually estimating the centroid and the separation of two incoherent point sources are maximally incompatible. Actually, reaching the quantum limit in one, inevitably, reduces the information of the other to zero~\cite{Lu:2021aa}.

However, this does not prevent one from a joint estimation of both quantities: it is possible to achieve acceptable performance in both variables without necessarily reaching the quantum limit in either. In~\cite{Shao:2022aa}, the authors numerically pursued optimal measures that approached the tradeoff relation and came remarkably close to achieving it. In the same vein, Ref.~\cite{Kimizu:2024aa} proposed an adaptive sequential protocol that uses optimal measurements (that have to be found numerically) to  simultaneously estimate both parameters.   

In this work, which serves as a seamless extension of preceding efforts, we analytically demonstrate the ideal measurement strategy as the sources get very close. This strategy hinges on a parameter whose manipulation enables us to glean enhanced insights from both centroid and separation metrics. In this way, we saturate the tradeoff relation and extract the maximum possible information from the two variables at the same time.  

This paper is organized as follows. In Sec.~\ref{sec:model}, we describe the mathematical formulation of our estimation problem. In Sec.~\ref{sec:optimal}, we derive our measurement scheme and demonstrate that the centroid and the separation can, in principle, be estimated simultaneously. Section~\ref{sec:Gaussian} studies the performance of our scheme for the relevant case of a Gaussian point spread function (PSF), showing that the measurements works  over a wide range of separations. Finally, our conclusions are summarized in Sec.~\ref{sec:conc}.

\section{Model and associated multiparameter Cram\'er-Rao bound}
\label{sec:model}

We  assume quasimonochromatic paraxial waves with one specified polarization and one spatial dimension, $x$ denoting the image-plane coordinate. For a linear spatially invariant system the corresponding object-plane coordinates can be obtained via the lateral magnification, which we take as unity without loss of generality.  

We phrase what follows in a quantum notation that will facilitate the calculations.  A wave of complex amplitude $U( x ) \in L^{2} (\mathbb{R})$ can then be assigned to a  ket $| U \rangle $, in such a way that $U( x )= \langle x | U \rangle$, with $| {x} \rangle$ representing an ideal point source localized exactly at ${x}$. 

The system is characterized by its PSF~\cite{Goodman:2004aa}, which describes the normalized intensity response to a point source.  We denote this PSF by $I (x) = | \langle x | \Psi \rangle |^{2} = |\Psi (x)|^{2}$, so that $\Psi(x)$ can be interpreted as the amplitude PSF.

Two point sources of equal intensities and separated by a distance $\mathfrak{s}$ are imaged by that system. Since they are incoherent with respect to each other, the total signal must be written as a density operator 
\begin{equation}
\label{eq:densmat}
  \varrho_{\bm{\mathfrak{s}}}= \tfrac{1}{2} ( \varrho_{+}  + \varrho_{-} ) \, ,
\end{equation}
where the total intensity is normalized to unity.  The individual
components $\varrho_{\pm}= |\Psi_{\pm}\rangle \langle\Psi_{\pm}|$ are
just $x$-displaced PSF states; that is, $ \langle x | \Psi_{\pm} \rangle = \langle x - \mathfrak{s}_{0} \mp \mathfrak{s}/2 | \Psi \rangle$, so that they are symmetrically located around the position of the geometric centroid~$\mathfrak{s}_{0}$. If we recall that the momentum operator (which in the $x$-representation acts as a derivative $P \mapsto - i \partial_{x}$)  generates displacements in the $x$ variable, we can write
\begin{equation}
 | \Psi_{\pm} \rangle = \exp[ - i (\mathfrak{s}_{0} \pm
 \mathfrak{s}/2) P ]  | \Psi \rangle \, .
\end{equation}
 
The density matrix \eqref{eq:densmat} depends on both the centroid $\mathfrak{s}_{0}$ and  the separation $\mathfrak{s}$. This is indicated by the vector $\bm{\mathfrak{s}} = (\mathfrak{s}_{0}, \mathfrak{s})^{\top}$. Our task is to estimate the values of $\bm{\mathbf{s}}$ from the outcomes  of some measurement performed on $\varrho_{\bm{\mathfrak{s}}}$.

Let us assume a real and symmetric PSF. This implies that $\bra{\Psi} P^{2n+1} \ket{\Psi}=0$ and we denote $\wp_{2n} =\bra{\Psi}P^{2n}\ket{\Psi}$ for $n=0,1,\ldots$ Our first step is to define an orthonormal computational basis by orthonormalizing the PSF state and its first and second derivatives, all centered at a good guess $\hat{\mathfrak{s}}_0$ for the centroid $\ket{\Psi_{\mathfrak{s}_0}} = \exp  (-i \hat{\mathfrak{s}}_0 \, P ) \ket{\Psi}$:  
\begin{equation}
\begin{split}
\ket{\Phi_1} & = \ket{\Psi_{\hat{\mathfrak{s}}_0}} \,, \\
\ket{\Phi_2} & = - \frac{i}{\sqrt{\wp_{2}}} \; P\ket{\Psi_{\hat{\mathfrak{s}}_0}}\,,\\
\ket{\Phi_3}& = \frac{1}{\sqrt{\wp_{4} - \wp_{2}^2}}(\wp_{2}- P^2)\ket{\Psi_{\hat{\mathfrak{s}}_0}}\,.
\end{split}
\end{equation} 

Next, we construct an optimal Positive Operator-Valued Measure (POVM)~\cite{Holevo:2003fv}. To this end, we introduce two vectors using a set of real-valued parameters $a_j$, $b_j$ and making one vector orthogonal to the signal PSF state: 
\begin{equation}
\label{POVM}
\begin{split}
\ket{\pi_1} & =  a_2 \ket{\Phi_2} + a_3 \ket{\Phi_3}, \\
& \\
\ket{\pi_2} & = b_1 \ket{\Phi_1} + b_2 \ket{\Phi_2}+b_3 \ket{\Phi_3} \, .
\end{split}  
\end{equation}
Then a family of three-element POVMs is designed as follows:
\begin{equation}
\Pi_1 = \ket{\pi_1}\bra{\pi_1} \, , 
\quad
\Pi_2 = \ket{\pi_2}\bra{\pi_2} \, , 
\quad
\Pi_3=1-\Pi_1-\Pi_2 \, .
\end{equation}  

Since our goal is to estimate small deviations of the signal state $\varrho_{\bm{\mathfrak{s}}}$ from the nominal PSF state $\ket{\Psi}\bra{\Psi}$, we  expand the signal state components in the small quantities $\delta_\pm=\mathfrak{x}_0\pm \mathfrak{s}/2$
\begin{equation}
\ket{\Psi_\pm} \approx \left(1-i\delta_\pm P-\frac{1}{2}\delta_\pm^2P^2+\frac{i}{6}\delta_\pm^3 P^3 \right)\ket{\Psi_{\hat{\mathfrak{s}}_0}}\, ,
\end{equation} 
where $\mathfrak{x}_0 = (\mathfrak{s}_0-\hat{\mathfrak{s}}_0)$ is the small deviation between the initial guess and the true centroid. This makes it possible to compute the probabilities $p(j | \bm{\mathfrak{s}}) = \Tr (\varrho_{\bm{\mathfrak{s}}} \Pi_{j})$ of observing $\Pi_j$ for true signal parameters $\bm{\mathfrak{s}}$. From these probabilities we can calculate the classical Fisher information (CFI) matrix associated to this measurement. The calculations are straightforward and we get  
\begin{equation}
\mathsf{F} (\bm{\mathfrak{s}} ) =  \left(
\begin{array}{cc}
\displaystyle 
4 \wp_{2} \,\left( \varepsilon_{\mathfrak{s}_0}^2 + \frac{\varepsilon_{\mathfrak{s}}^2}{1+\mathfrak{r}} \right) + \mathcal{O}(\mathfrak{s}^2) &  
\displaystyle \frac{4\varepsilon_{\mathfrak{s}}^2}{4\mathfrak{r}+1/\mathfrak{r}} + \mathcal{O}(\mathfrak{s}) \\[4mm]
\displaystyle 
\frac{4\varepsilon_{\mathfrak{s}}^2}{4\mathfrak{r}+1/\mathfrak{r}} + \mathcal{O}(\mathfrak{s}) & \displaystyle  \wp_{2} \,\varepsilon_\mathfrak{s}^2\frac{\mathfrak{r}}{1+\mathfrak{r}} + O(\mathfrak{s}^2)
\end{array}
\right)\,,
\end{equation}
where we have enforced $a_2,b_1,b_2>0$ and the parameters
\begin{equation}
\label{deltas}
\varepsilon_\mathfrak{s}^2=a_2^2\, ,\quad  \varepsilon_{\mathfrak{s}_0}^2=\frac{b_2^2}{1-b_1^2}\, ,\quad 
\varepsilon_\mathfrak{s}^2,\, \varepsilon_{\mathfrak{s}_0}^2 \le 1 \, .
\end{equation} 
We also set $\mathfrak{r} = \tfrac{1}{4} (\mathfrak{s}/\mathfrak{x})^2$ as the ratio of the separation to the prior knowledge of the centroid. Similar to the intuition from postselected metrology~\cite{Jenne:2022aa}, quantum measurements effectively allow one to zoom in on a parameter range and there obtain enhanced sensitivity; this means, to harness the advantages, one must have prior information about the region to magnify, so we henceforth consider $\mathfrak{r}\gg 1$. This allows us to consider the CCRB
\begin{equation}
\mathsf{C}(\hat{\bm{\mathfrak{s}}}) \succcurlyeq \mathsf{F}^{-1} (\bm{\mathfrak{s}}) \, ,
\end{equation} 
where the matrix inequality $\mathsf{A} \succcurlyeq \mathsf{B}$ means that $\mathsf{A} - \mathsf{B}$ is a positive semidefinite matrix.  Here, $\mathsf{C}_\Psi (\hat{\bm{\mathfrak{s}}})$ is the covariance matrix of the unbiased estimator $\hat{\bm{\mathfrak{s}}}$. 

In quantum estimation theory, we often consider the QFI matrix  instead of the classical version to evaluate the error bound. It is known that the $\mathsf{F} ( \bm{\mathfrak{s}} )$ for a given measurement scheme is bounded from above by the QFI matrix $ \mathsf{Q} (\bm{\mathfrak{s}})$; that is
\begin{equation}
\label{eq:min}
\mathsf{F} (\bm{\mathfrak{s}}) \preccurlyeq \mathsf{Q} (\bm{\mathfrak{s}}) \, .
\end{equation}
If there exists a measurement that achieves the upper bound in \eqref{eq:min}, it is the optimal measurement. As a rank-2 state~\cite{Toth:2018aa}, elements of the QFI can be found analytically from the purity of the state and the variance of $P$, giving
\begin{equation}
\mathsf{Q}_{\mathfrak{s}_0\mathfrak{s}_0}=4[\wp_2 -|\langle\Psi_+|\Psi_-\rangle|^2(1-|\langle\Psi_+|\Psi_-\rangle|^2)/4]\approx 4\wp_2(1-\mathfrak{s}^2/4) \, .
\end{equation}
By convexity properties of the QFI and its further relation to variances~\cite{Toth:2022aa}, we can immediately bound $\mathsf{Q}_{\mathfrak{s}\mathfrak{s}}\leq \wp_2$. In our case, since our CFI matrix is diagonal as $\mathfrak{r} \rightarrow \infty$ and $\mathfrak{s}\rightarrow 0$, we have  
\begin{equation}
\mathsf{F}_{jj} = \varepsilon_j^2 \mathsf{Q}_{jj}
\end{equation}
where the subscript $j$ runs the values $\mathfrak{s}$ and $\mathfrak{s}_{0}$. The same applies to the inverse diagonal elements of the corresponding inverse matrices, which set the classical and quantum bounds. 

\section{Optimal measurement}
\label{sec:optimal}

Our main results is that any three-element POVM of the form \eqref{POVM} with $a_2,b_1,b_2>0$ generates, in the limit of small separations, information about centroid and separation, which differ from the corresponding quantum bounds by constant factors. This amounts to lifting the Rayleigh curse. 

\begin{figure}[b]
  \includegraphics[width=0.80\columnwidth]{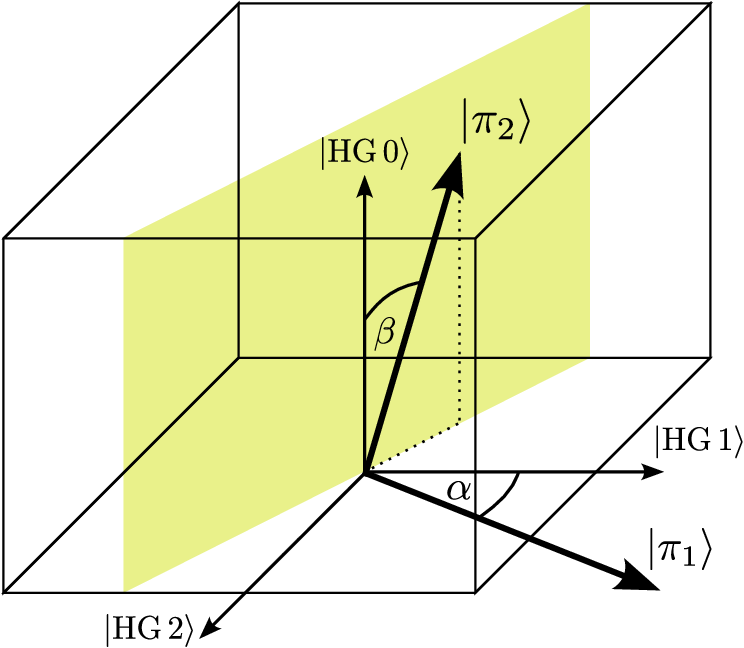}
  \caption{The geometry of optimal measurements. Aligning modes $\ket{\pi_1}$ or $\ket{\pi_2}$ along any coordinate axis must be avoided to not invalidate the optimality conditions. For non Gaussian PSFs the HG modes are replaced with the corresponding three lowest-order orthonormalized PSF derivatives.  \label{figpovm}}
\end{figure}

The next step is optimizing POVM with respect to $a_j$ and $b_j$. Unfortunately, the obvious choice $a_2=1$ and $b_1^2=1-b_2^2$, which (apparently) saturates the QCRB fails, as it is not consistent with $\Pi_3$ being positive semidefinite. Considering the projector $A=\ket{\Phi_1}\bra{\Phi_1}+\ket{\Phi_2}\bra{\Phi_2}$ and imposing $\Pi_3\ge 0$ on this subspace gives
\begin{equation}
1-\det(A\, \Pi_3 \, A) = a_2^2(1-b_1^2)+b_1^2+b_2^2\le 1.
\end{equation}
Using Eq.~\eqref{deltas} and rearranging, we get the trade-off relation 
\begin{equation}
\label{tradeoff}
\varepsilon_\mathfrak{s}^2 + \varepsilon_{\mathfrak{s}_0}^2\le 1
\end{equation}
that places limits on how the total information can be distributed between the centroid and separation variables. 
For example, letting $\varepsilon_\mathfrak{s}^2=\varepsilon_{\mathfrak{s}_0}^2\le 1/2$, the price to pay for having a ``uniform'' access to both centroid and separation in a multiparameter estimation amounts to minimally doubling their mean square errors with respect to single-parameter cases. This is consistent with the tradeoff relation introduced in \cite{Lu:2021aa} for the so-called information regrets, defined as
\begin{equation}
\Delta_j = \sqrt{\frac{\mathsf{Q}_{jj}- \mathsf{F}_{jj}}{\mathsf{Q}_{jj}}}
\end{equation}
which in the limit of small separations take the form~\cite{Shao:2022aa}
\begin{equation}
	\Delta_{\mathfrak{s}}^2+\Delta_{\mathfrak{s}_{0}}^2 \geq	1 \, .
	\label{eq:tradeoff_Shao}
\end{equation}

\begin{figure}
  \includegraphics[width=0.90\columnwidth]{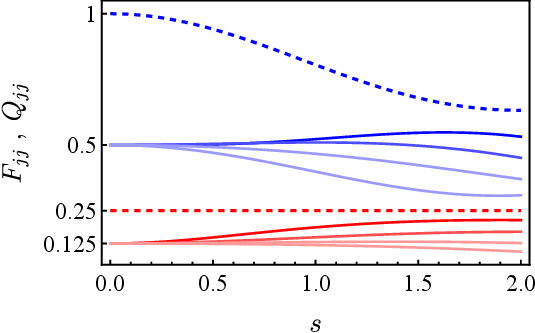}
  \caption{The diagonal entries of the CFI (blue) and of the QFI (red) matrices for the centroid (dashed) and the separation (solid) variables. The decreasing values are $\beta=\pi/12,\pi/6,\pi/4,\pi/3$. 
  \label{figperform}}
\end{figure}

For our POVM, we have $\Delta_j^2=1-\varepsilon_j^2$, and in the limit of small separation, ~\eqref{tradeoff} is equivalent to \eqref{eq:tradeoff_Shao}.

One particular way of saturating those trade-offs is by choosing
\begin{equation}
\ket{\pi_1}=  \begin{pmatrix}
0\\ 
\cos \alpha\\
\sin \alpha 
\end{pmatrix} \, ,
\qquad
\ket{\pi_2}= 
\begin{pmatrix} 
\cos \beta \\
\sin \alpha \sin \beta \\
-\cos\alpha \sin \beta 
\end{pmatrix}
\end{equation}
so that $\varepsilon_\mathfrak{s}^2 = \cos^{2} \alpha$ and $\varepsilon_{\mathfrak{s}_0}^2 = \sin^{2} \alpha$, ensuring we have explicitly found the best measurement possible for very small separations—the only interesting regime. 

Visualizing $\ket{\pi_1}$ and $\ket{\pi_2}$ as vectors in a real 3D space, whose $z$, $x$, and $y$-coordinates (in this order) are given by the coefficients $a_\ell$ and $b_\ell$ $( \ell \in \{ 1, 2, 3\} )$ as in Eq.~\eqref{POVM}, the states on which to project are as follows: $\ket{\pi_1}$ lies in the $xy$ plane making an angle $\alpha$ with the $x$-axis. This angle $\alpha$ determines how the information is distributed between the centroid and the separation estimates according to \eqref{tradeoff}. The second projection lies in a plane perpendicular to $\ket{\pi_1}$, while its orientation in this plane, i.e. angle $\beta$, can be chosen arbitrarily. This is indicated in Fig.~\ref{figpovm}. In particular, rotating the POVM so as to move $\ket{\pi_1}$ closer to $\ket{\Phi_1}$ (the first PSF derivative) improves the estimation of separation while moving it closer to $\ket{\Phi_2}$ (the second PSF derivative) improves the estimation of centroid. In both cases this is at the expense of worsening the estimates of the other variable.

\section{Gaussian PSF}
\label{sec:Gaussian}

The freedom of rotating $\ket{\pi_2}$ about $\ket{\pi_1}$ by (almost) any angle $\beta$ without compromising the small-separation performance of the optimal POVM is a useful feature. First, projections on some sets of modes may offer easier or more robust implementation in a laboratory and some POVM orientations may be preferred over others for this reason. Second, different settings of $\beta$ alter the higher-order terms of the CFI matrix expansion. This becomes important when extending the range of metrology to larger separations. 

This will be demonstrated with a conceptually simple yet practically important case of a Gaussian PSF  of variance $\sigma$:
\begin{equation}
\Psi (x) = \braket{x|\Psi}= \frac{1}{(2\pi \sigma^{2})^{\tfrac{1}{4}}} 
\exp \left ( - \frac{x^2}{4\sigma^{2}} \right ) \, ,
\end{equation}
Here, the computational basis $\ket{\Phi_j}$ becomes the set of HG modes and the POVM elements, as labeled in Fig.~\ref{figpovm}.

\begin{figure}[t]
	\includegraphics[width=0.90\columnwidth]{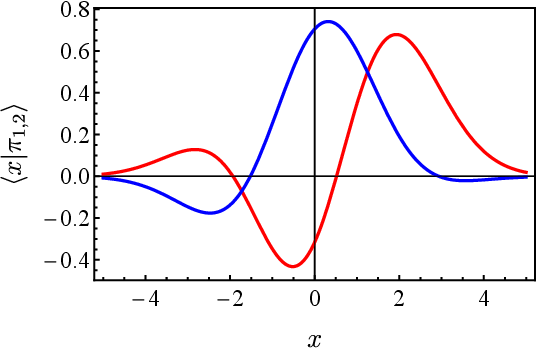}
	\caption{Optimal measurement modes $\ket{\pi_1}$ (red) and $\ket{\pi_2}$ (blue) with $\alpha=\pi/4$ and $\beta=\pi/6$.
  \label{figwave}}
\end{figure}

Going for balanced resolutions in both variables, we set $\alpha=\pi/4$ and compare the performances of POVMs with different $\beta$ settings to their respective quantum limits. The results appear in Fig.~\ref{figperform}. For very small separations, the $\beta$ setting is irrelevant, as we already concluded. Making the separation larger gives clear advantage to small $\beta$ values, for which the resolution in separation at actually gets close to the quantum limit at some points. Figure~\ref{figwave} provides a visualization of one pair of such optimal modes. There is no point in extending the analysis to even larger separations, because there the two signal components spatially separate, Rayleigh's curse does not apply, and direct intensity detection starts to rule. 

\section{Concluding remarks}
\label{sec:conc}

We have introduced an optimal measurement, valid for an arbitrary PSF, that allows extracting the maximum possible information from the centroid and the separation of two incoherent point sources when they are very close. This measurement can be adjusted at will, allowing us to extract more information from the centroid or the separation, ensuring that it always extracts the maximum total information. We have also demonstrated, for a Gaussian PSF, the good performance of this measure for small and moderate separations.

\acknowledgments
We acknowledge discussions with M.~Tsang at an early stage of this work. This publication has received financial support from the QuantEra program (project ApresSF) and from the Agencia Estatal de Investigaci\'on (Grant PID2021-127781NB-I00). A. Z. G. acknowledges that the NRC headquarters is located on the traditional unceded territory of the Algonquin Anishinaabe and Mohawk people, as well as support from the NSERC PDF program and the NRC Quantum Sensors Challenge program. L. L. S. S. was supported in part by the grant NSF PHY-1748958 to the Kavli Institute for Theoretical Physics (KITP).


%

\end{document}